\documentclass[10pt]{article}
\usepackage{amsmath}
\usepackage{amsfonts}
\usepackage{multicol}
\usepackage{fancyhdr}
\usepackage{graphicx}
\usepackage[text={7.5in,9.5in},nohead,centering]{geometry}

\newcommand{\iuparrow}{\smash{\lower .5ex \hbox{$\uparrow$}}}
\newcommand{\idownarrow}{\smash{\raise .5ex \hbox{$\downarrow$}}}

\newcommand{\inearrow}{\smash{\lower .7ex \hbox{$\!\nearrow$}}}
\newcommand{\isearrow}{\smash{\raise .7ex \hbox{$\!\searrow$}}}
\newcommand{\inwarrow}{\smash{\lower .7ex \hbox{$\nwarrow\!$}}}
\newcommand{\iswarrow}{\smash{\raise .7ex \hbox{$\swarrow\!$}}}

\begin{document}

\def\<{\langle}
\def\>{\rangle}
\def\be{\begin{equation}}
\def\ee{\end{equation}}
\def\bea{\begin{eqnarray}}
\def\eea{\end{eqnarray}}
\def\qed{\leavevmode\unskip\penalty9999 \hbox{}\nobreak\hfill
     \quad$\blacksquare$
     \par\vskip3pt}
\def\Pr{\textrm{Pr}}
\def\h{\frac{1}{2}}
\def\hh{\tfrac{1}{2}}

\def\tbit#1#2#3#4{\begin{array}{|c|c|c|c|}\hline #1&#2&#3&#4
\\\hline\end{array}}

\clearpage

\begin{center}
\textbf{How Einstein and/or Schr\"odinger should have discovered Bell's Theorem in 1936}
\bigskip

Terry Rudolph

Department of Physics, Imperial College London, Prince Consort Rd, London SW7 2AZ

\bigskip
\today
\end{center}

\begin{quotation}

This note shows how one can be led from considerations of quantum steering to Bell's theorem. The point is that steering remote systems by choosing between two measurements can be described in a local theory if we take quantum states to be associated many-to-one with the underlying "real states" of the world. Once one adds a third measurement this is no longer possible. Historically this is not how Bell's theorem arose - there are slight and subtle differences in the arguments - but it could have been.

\end{quotation}

\bigskip
\bigskip
\textbf{AFTERWORD}

Following is the appendix of an incomplete paper from mid-2003\footnote{The only changes I have made are to add references and change 1932 to 1936!}, that I completely forgot existed until a bit over a year ago when a very nice talk by Howard Wiseman\footnote{Based primarily around H. Wiseman, Contemporary Physics {\bf 47}, 79-88 (2006); quant-ph/0509061} triggered me into searching through old notes for my vaguely recollected version of ``Bell's theorem via steering''. The somewhat long full paper titled \textit{Quassical Mechanics} is incomplete, it primarily contains a variety of examples of ``toy theories'' following the ideas of Rob Spekkens\footnote{R.W. Spekkens, Phys. Rev. A 75, 032110 (2007); quant-ph/0401052}. One of them (eventually!) led to, and was superseded by, arxiv:1111.5057. Having given up on myself getting around to completing it anytime soon, but having had a discussion with Reinhard Werner the week before last during which he expressed the opinion that `Einstein should have discovered Bell's theorem via steering', I'm posting this particular part of it as-is. The simple structure of the argument has not quite been captured yet by recent work on steering and nonlocality (primarily by Wiseman and colleagues\footnote{H. M. Wiseman, S. J. Jones, and A. C. Doherty, Phys. Rev. Lett. 98, 140402 (2007)}).

Basically the appendix is about how, what we would now call a ``$\psi$-epistemic'' interpretation of quantum states (following Harrigan and Spekkens\footnote{N. Harrigan, R.W. Spekkens, Found. Phys. 40, 125 (2010);quant-ph/0706.2661.}), can be used to save locality when one considers steering the remote quantum state of a system using only \emph{two} measurements, as was done in the EPR paper. However, as soon as one adds a generic third measurement, locality cannot be saved. This seems to contradict the well known fact that CHSH inequality violation only requires a choice between two measurements. But that argument actually relies on looking at correlations of the two measurements with a pair of measurements at the remote system as well. The argument I'm interested in here is about what just one party can infer about the ``real state'' of affairs at the remote system necessarily being changed (``steered'') nonlocally based solely on their ability to steer its quantum state by, in this case, one of \emph{three} different measurements. Why I like it is that one never talks about the real state of affairs (the ``ontic state'') of the system being measured to do the steering.

The origins of my thinking at all about classical versus quantum steering go back to working with Rob Spekkens on two party cryptography\footnote{R.W. Spekkens and T. Rudolph, Quantum Inform. Compu. 2, 66 (2002); quant-ph/0107042.}${}^,$\footnote{T. Rudolph and R.W. Spekkens, Phys. Rev. A 70, 052306 (2004); quant-ph/0310060} where steering plays a crucial role, and much of my thinking was influenced of course by discussions with him. After he sent me his first ideas about and proofs of preparation contextuality\footnote{R.W. Spekkens, Phys. Rev. A 71, 052108 (2005); quant-ph/0406166} I simplified them based around what I knew from this simple nonlocality proof, and conversely in this version below I mention preparation non-contextuality as being the constraint that locality imposes for this style of argument. However; as far as I can see the precise and interesting connections between proofs of nonlocality and proofs of preparation contextuality have still not been completely fleshed out, though Barrett (private communication) has made some progress in this regard.

\clearpage

\textbf{APPENDIX A of incomplete article \textit{Quassical Mechanics} - draft of July 29, 2003}

\begin{multicols}{2}
\begin{appendix}
\section{\large How Einstein and/or Schr\"odinger should have discovered
Bell's theorem in 1936}

Bell's theorem - the empirical fact that features of this universe
cannot be described by a local theory - is a statement of physics
which transcends merely quantum mechanics. Bell's theorem is the
only facet of quantum mechanics I believe will still be considered
a fascinating insight into nature in a few hundred years time.

In this appendix I will attempt a little revisionist history. In
particular, I will attempt to show how a very simple argument
establishing the impossibility of a local hidden variable (LHV)
description of QM was lingering on the edge of Schr\"odinger's and
Einstein's consciousness in 1936. In particular, in 1936 the two
of them, via various correspondences [1], were collectively
considering the following features of QM:
\begin{itemize}
\item \emph{The quantum mechanical wavefunction may not be a
complete description}. The possibility that the wavefunction was
an epistemic ``catalog of information'' was under consideration.
\item \emph{The possibility of steering.} Inspired by the EPR paper,
Schr\"odinger had proven the quantum steering theorem, in large
(though not complete) generality.
\end{itemize}

Both knew that if pure quantum states are taken to be states of
reality, then the possibility of steering is violently
incompatible with locality. In fact, the term `steering' was
chosen by Schr\"odinger precisely to reflect this fact - in such
scenarios it seems that an action performed on one half of an
entangled system nonlocally ``steers'' or ``drives'' the
wavefunction of the other system\footnote{At the end of his paper
on steering Schr\"odinger mused that perhaps the resolution would
be found in a certain dephasing process (known today as
`decoherence') which prevents us from creating spatially separated
entangled states in practise. This has turned out not to be the
case.}. The purpose of this section is to show how, by a simple
argument, this conceptual incompatibility could have been proven
algebraically to hold for all LHV theories, thereby establishing
what we know today as Bell's theorem.

The quantum steering theorem is [2]:\\
\emph{\textbf{Theorem}: Given an entangled state $|\psi_{AB}\>$ of two
systems $A,B$, a measurement on system $A$ can collapse system $B$
to the set of states $\{|\phi_i\>\}$ with associated probabilities
$p_i$, if and only if
\[
\rho_B=\sum_i p_i |\phi_i\>\<\phi_i|,
\]
where $\rho_B\equiv Tr_A|\psi_{AB}\>\<\psi_{AB}|$ is the reduced
state of system $B$.}\\
Schr\"odinger in fact only proved the theorem for ensembles of
states $|\phi_i\>$ which are linearly independent (possibly
non-orthogonal); this is more than we will need here.

In examining the description of steering in a local hidden
variable theory, we presume that the actual physical properties of
system $B$ are described by a complete set of variables $\lambda$.
No claims are made about the specific nature of these variables,
other than they should correctly reproduce the predictions of QM.
This entails certain restrictions. For instance, consider a
von-Neumann measurement described in QM by the two projection
operators $|\chi\>\<\chi|,I-|\chi\>\<\chi|$. We know that there is
a state $|\chi\>$ of the system,  which gives one measurement
outcome with certainty, the other with probability zero. Since the
measurement outcomes are presumed to be dictated by the particular
value of $\lambda$ governing the physics of the system, we see
that the set of all possible $\lambda$ for the system contains at
least two disjoint sets - a set of those values which yield
outcome $|\chi\>\<\chi|$ with certainty and those which yield
$I-|\chi\>\<\chi|$ with certainty. (There could in general be
values of $\lambda$ which lead to neither outcome with certainty).
We denote by $S_\chi$ the subset of $\lambda$ values which lead to
outcome $\chi$ with probability 1.

In a steering scenario, system $B$ is described quantum
mechanically by the mixed state $\rho_B$. We know that this state
can be steered to the eigenstates of $\rho_B$, which are
orthogonal. Since each of these eigenstates are associated with
disjoint values of $\lambda$, we see that, under a presumption of
locality, $\rho_B$ must be associated with a probabilistic
distribution over at least two different $\lambda$. We denote the
set of all $\lambda$ underlying $\rho_B$ by $S_\rho$, and denote
by $\nu(\lambda)$ any distribution over $S_\rho$ that is the
`hidden variable' description of $B$. The presumption of locality
also indicates that a measurement on system $A$ cannot change the
`real state of affairs' at $B$ - in particular, therefore, it
cannot change the value of $\lambda$ governing $B$, and thus
$\nu(\lambda)$, which is used by the observer at $B$ to describe
their system, is unaffected by the measurement performed at $A$.
For simplicity, from here on we limit ourselves to the case where
$\rho_B$ two-dimensional, and further we will take $\rho_B=I/2$,
that is, the maximally mixed state.

Let us first formalize the reasoning of Schr\"odinger and
Einstein, which yields a simple argument against local hidden
variables if pure quantum states are `states of reality'. More
precisely, we examine the possibility that pure states are ontic -
they correspond to a definite value of $\lambda$, while mixed
quantum states are epistemic - they correspond to a distribution
over some $\lambda$. Thus, in the ontic view, the state $|x\>$
actually corresponds to some specific value $\lambda_x\in S_x$, we
therefore associate $|x\>$ with a delta function distribution
$\delta(\lambda_x)$ over the hidden variables. We need only
consider the case where steering is performed either to a pair of
orthogonal states $|x\>,|X\>$ or to another pair of orthogonal
states $|y\>,|Y\>$, with $0<|\<x|y\>|^2<1$. That is,
\[
\rho_B=\hh|x\>\<x|+\hh|X\>\<X|=\hh|y\>\<y|+\hh|Y\>\<Y|.
\]
Locality ensures that $S_{\rho_B}=S_x \cup S_X=S_y\cup S_Y$. (Thus
all values of $\lambda\in S_\rho$ would yield one of the
measurement outcomes $|x\>\<x|, |X\>\<X|, |y\>\<y|$ or $|Y\>\<Y|$
with certainty.)  However, the crucial use of locality is to
enforce \emph{preparation non-contextuality}
[3]. That is, regardless of questions of locality,
in order for an ontic interpretation of pure states to be
consistent, it is necessary that two different preparation
procedures leading to the same mixed state are actually described
by different distributions over the hidden variables. For example,
in this case, one needs that $\hh
\delta(\lambda_x)+\hh
\delta(\lambda_X)=\nu_1(\lambda)$, while $\delta(\lambda_y)+\hh
\delta(\lambda_Y)=\nu_2(\lambda)$, where the two distributions
$\nu_1(\lambda),\nu_2(\lambda)$ are both valid hidden variable
descriptions of $\rho_B$. This requirement shows that the
procedure for preparing $\rho_B$ is necessarily contextual in such
an interpretation. In the steering scenario, the initial
distribution $\nu(\lambda)$ is unaffected by the measurement at
$A$. Hence the role of locality is to enforce $\nu_1=\nu_2$, which
then implies
\[
\nu(\lambda)=\hh \delta(\lambda_x)+\hh \delta(\lambda_X)=\hh \delta(\lambda_y)+\hh \delta(\lambda_Y).
\]
Such a description is inconsistent, by virtue of the fact that
within the ontic view we necessarily have
$\lambda_{x}\neq\lambda_{X}\neq
\lambda_{y}\neq\lambda_{Y}$. Such an argument contains the essence of what
disturbed Einstein and Schr\"odinger, in a slightly complicated
form.
%

If pure quantum states are epistemic, however, we must go a little
further in order to rule out local hidden variable theories. Under
the epistemic view, the process of steering is simply reflects the
change in information that the observer holding system $A$ has
about the system $B$, based upon their measurement outcome on
system $A$. The particular correlation between $A$ and $B$ is
presumed known of course. As we have seen, steering appears in
some form both classically and quassically, which are local
physical descriptions. Quassically we even obtain steering to
multiple different pure state decompositions. However, the
argument below shows that quassically we cannot simulate all such
steering scenarios.

Let us use the notation that $x(\lambda)$ denotes the distribution
over $S_x$ corresponding to the state $|x\>$. As mentioned,
locality ensures that the distribution $\nu(\lambda)$ is not
affected by the measurement at $A$. Clearly we must have
\be\label{nu}
\nu(\lambda)=\hh x(\lambda)+\hh X(\lambda)=\hh y(\lambda)+\hh Y(\lambda)
\ee
Normalization relations of the form $\int_{S_{x}}\!\!d\lambda\,
x(\lambda)$ must be satisfied. The distributions $x(\lambda),
y(\lambda)$ cannot be disjoint (since if $S_y\subset S_X$ then the
probability of obtaining an outcome $|x\>\<x|$ when a system is in
the state $|y\>$ would be zero). That is, there is an overlap
between the regions $S_x$ and $S_y$, which we denote $S_{1}$. Note
that values of $\lambda$ in this region yield measurement outcomes
$|x\>\<x|$ and $|y\>\<y|$ with certainty - the non-orthogonality
of $|x\>,|y\>$ is reflected in the fact that the distribution
$y(\lambda)$ only partially overlaps $S_x$. More precisely, in
order to conform with the predictions of QM, we must have that
\be\label{int}
\int_{S_{x}}\!\!d\lambda\;y(\lambda)=\int_{S_{1}}\!\!d\lambda\;y(\lambda)=
|\<x|y\>|^2\equiv\alpha.
\ee
In fact there are 4 disjoint regions of the $\lambda$-space to
consider: $S_1\equiv S_x \cap S_y$, $S_2\equiv S_x \cap S_Y$,
$S_3\equiv S_X
\cap S_y$, $S_4\equiv S_X \cap S_Y$. We will use the notation that
\[
x_j\equiv \int_{S_{j}}\!\!d\lambda\;x(\lambda),\;\;\; j=1,\ldots,4
\]
and so on.

Clearly, by integrating (\ref{nu}) over the appropriate regions,
we have the following constraints:
\be\label{12}
\nu_j = \hh x_j+\hh X_j=\hh y_j+\hh Y_j, \;\;\; j=1,\ldots,4.
\ee
From equations of the form (\ref{int}) we obtain
\begin{eqnarray*}
&x_1=y_1=X_4=Y_4=\alpha& \\
&x_2=y_3=X_3=Y_2=1-\alpha&
\end{eqnarray*}
with all other values equal to 0. Thus, by (\ref{12}),
$\nu_1=\nu_4=\alpha/2$, while $\nu_2=\nu_3=(1-\alpha)/2$.

In order to obtain a contradiction, we need to consider a third
pair of orthogonal states $|z\>,|Z\>$ which, by the steering
theorem, can also be steered to via a measurement on $A$. For
simplicity, we presume that the state $|z\>$ `bisects' (has equal
overlap with) the states $|x\>,|y\>$. Thus
\[
|\<z|x\>|^2=|\<z|y\>|^2=|\<Z|X\>|^2=|\<Z|Y\>|^2\equiv
\beta=\hh(1+\sqrt{\alpha}),
\]
the last term being the quantum mechanical prediction. From this
we deduce that
\bea
&z_1+z_2=\beta=z_1+z_3,\; z_3+z_4=1-\beta=z_2+z_4,&\\
&Z_3+Z_4=\beta=Z_2+Z_4,\; Z_1+Z_2=1-\beta=Z_1+Z_3.&
\eea
Clearly $z_2=z_3$ and $Z_2=Z_3$. We must also have
\[
\nu_j=\hh z_j+\hh Z_j \;\;\; j=1,\ldots,4.
\]

There is no way to satisfy all these equations, subject to the
necessary requirement $z_j,Z_j\ge 0$. For example, an independent
set of the above equations is
\bea
z_1+z_2=Z_2+Z_4&=&\beta \label{aa}\\
z_2+z_4=Z_1+Z_2&=&1-\beta \label{bb}\\
z_1+Z_1&=&\alpha \label{cc}
\eea
From these we get
\[Z_1=\alpha-z_1=\alpha-(\beta-z_2)=\alpha-\beta+(1-\beta-z_4)=1-2\beta+\alpha-z_4,\]
which, using $\beta=\hh(1+\sqrt{\alpha})$, gives
$Z_1=\alpha-\sqrt{\alpha}-z_4$. This is manifestly negative for
any $0\le \alpha,z_4\le1$. This completes the demonstration of
incompatibility between local realism and QM.

Although this proof is algebraic and thus reminiscent of GHZ type
proofs against local realism, it is in fact more or less
equivalent to Mermin's exposition of Bell inequalities in
[4].

\textbf{REFERENCES}

[1] A. Einstein, Letter to Schr\"odinger (1935). Translation
from D. Howard, Stud. Hist. Phil. Sci., 16, 171 (1985).

[2] E. Schr\"odinger. Proc. Camb. Phil. Soc., 31, 555 (1935);
Proc. Camb. Phil. Soc., 32, 446 (1936).

[3] R. W. Spekkens, Phys. Rev. A, 71, 052108 (2005),
arXiv:quant-ph/0406166.

[4] N.D. Mermin, Am. J. Phys \textbf{49}, 940 (1981)

\end{appendix}

\end{multicols}

\end{document}